\documentclass[twocolumn,showpacs,prl,aps]{revtex4-1}
\usepackage{bm,color,amsmath,amssymb,mathrsfs,dsfont,latexsym,graphicx,psfrag}

\newcommand{\bs}[1]{\boldsymbol{#1}}
\newcommand{\be}{\begin{equation}}
\newcommand{\ee}{\end{equation}}
\newcommand{\bea}{\begin{eqnarray}}
\newcommand{\eea}{\end{eqnarray}}

\renewcommand{\phi}{\varphi}
\renewcommand{\epsilon}{\varepsilon}

\begin{document}

\title{Reply to comment on "Unconventional Fermi Surface
  Instabilities in the Kagome Hubbard Model'' by Kiesel et al., Phys. Rev. Lett. 110, 126405 (2013)}
\author{Ronny Thomale} 
\email{rthomale@physik.uni-wuerzburg.de}
\affiliation{Institut f\"ur Theoretische Physik, Universit\"at W\"urzburg,
  Am Hubland, D-97074 W\"urzburg, Germany}

\date{\today}

\begin{abstract}
We respond to the comment arXiv:1809.03931 put forward by Li-Han Chen,
Zhen Liu, and Jian-Tin Zheng on our work Kiesel, Platt, and Thomale, Phys. Rev. Lett. {\bf 110}, 126405
(2013). All misunderstandings and misconceptions communicated in the
comment are addressed in light of the status quo of functional
renormalization group.   
\end{abstract}
\maketitle

In a recent comment~\cite{comment}, concerns are raised on the
validity of our results
obtained for the kagome Hubbard model with
functional renormalization
group~\cite{PhysRevLett.110.126405}. Specifically, the
criticism applies to a local Hubbard model of coupling strength $U$ in the intermediate
to strong coupling limit, and a nearest
neighbor kagome hopping model tuned to van Hove filling. Employing Hartree
Fock mean field theory, dynamical mean field theory (DMFT), and variational
quantum Monte Carlo (VQMC) all details of
which are concealed, the authors purportedly obtain the following
results: (i) The bare susceptibility implies a preference to
ferromagnetism below $U_0\approx 1.57$ and
120$^\circ$ N{\'e}el order above it (Fig. 1a in~\cite{comment}); (ii)
the DMFT finds a switch from a dominant ferromagnetic moment to a
120$^\circ$ moment at around the same $U_0$ (Fig. 1b
in~\cite{comment}); (iii) as a function of undisclosed variational
parameters on a 108 site lattice with 90 electrons, the
authors find that among the tested ordering patterns of charge density
wave (unspecified), 120$^\circ$ N{\'e}el order, and a 12 site unit
cell "David-star'' pattern, the latter prevails upon VQMC
optimization (Fig. 1c in~\cite{comment}). 

It is not our intent to reveal the inherent inconsistencies of this
comment itself, but only address it to the extent that it concerns our
work~\cite{PhysRevLett.110.126405}. Therein, we have reported the
$U$-$V$ Hubbard phase diagram, where $V$ denotes the nearest-neighbor
interaction, within a 1-loop functional renormalization group
treatment~\cite{RevModPhys.84.299,doi:10.1080/00018732.2013.862020}. For dominant $U$, both at and in the vicinity of
van Hove filling, we find ferromagnetic fluctuations to dominate even
up to values of $U=10$, in contrast to $U_0\approx 1.57$ stated
in~\cite{comment}. By that, we mean that as a function of the
FRG cutoff parameter, the RG flow looks such that the ferromagnetic
channel, i.e., the $S=1$ particle-hole condensate with ${\bs q}=0$,
dominates as the 2-particle vertex flows to strong coupling. Our numerical finding is supported by the analytical
result obtained by us in the weak coupling limit, where we have
discovered the sublattice interference mechanism leading to a
suppression of the $U$ matrix elements for nested momentum scattering~\cite{PhysRevB.86.121105}.

As expected for a subtle physical problem such as the kagome Hubbard
model, results obtained by alternative numerical methods show large agreement, but also partial
disagreement. To single out the most relevant one so
far reported for the kagome Hubbard model in the itinerant regime,
Ref.~\cite{PhysRevB.87.115135} employs singular mode functional
renormalization group; it confirms the pivotal relevance of the
sublattice interference mechanism~\cite{PhysRevB.86.121105}, as well as its natural tendency to
promote ${\bs q}=0$ instabilities. As mentioned both
in~\cite{PhysRevLett.110.126405} and~\cite{PhysRevB.87.115135}, aspects
such as Fermi surface resolution as well as the precise treatment of the
kagome flat band do matter for larger interaction strength, and remain
subject to future research and methodological reconciliation. In
addition, since the FRG approach eo ipso assumes electronic itineracy and, as
any other diagrammatic approach for interacting electrons such as
random phase approximation, is
only strictly controlled in the weak coupling limit, the precise
convergence radius as a function of interaction strength is
notoriously difficult to determine, notwithstanding promising recent efforts of
methodological refinement~\cite{PhysRevLett.120.057403}.

Independent of those aspects that are typical to any academic
development of a numerical method, we allow ourselves the assertion 
that the results in Ref.~\cite{PhysRevLett.110.126405} are certainly far from
"spurious'', as claimed in~\cite{comment}, but constitute fundamental insights into exotic Fermi
surface instabilities found in the kagome Hubbard model, where they
conspire with the sublattice interference mechanism~\cite{PhysRevB.86.121105}. If
comment~\cite{comment} should have any constructive impact, it will
hopefully stimulate the ongoing search for and analysis of correlated kagome metals~\cite{kmetal}.


\end{document}